\documentstyle[11pt,amssymb,epsf]{article}

\makeatletter
\@addtoreset{equation}{section}
\makeatother

\def\ben{\begin{equation}}
\def\een{\end{equation}}

\let\a=\alpha    
    
 \let\m=\mu \let\n=\nu

\let\C=\Chi

\def\nn{\nonumber} \def\bd{\begin{document}} \def\ed{\end{document}}
\def\ds{\documentstyle} \let\fr=\frac \let\bl=\bigl \let\br=\bigr
\let\Br=\Bigr \let\Bl=\Bigl
\let\bm=\bibitem
\let\na=\nabla
\let\pa=\partial \let\ov=\overline
\newcommand{\be}{\begin{equation}}
\newcommand{\ee}{\end{equation}}
\def\ba{\begin{array}}
\def\ea{\end{array}}
\def\ft#1#2{{\textstyle{{\scriptstyle #1}\over {\scriptstyle #2}}}}
\def\fft#1#2{{#1 \over #2}}
\def\del{\partial}
\def\vp{\varphi}
\def\sst#1{{\scriptscriptstyle #1}}
\def\oneone{\rlap 1\mkern4mu{\rm l}}
\def\td{\tilde}
\def\wtd{\widetilde}
\def\ie{\rm i.e.\ }
\def\dalemb#1#2{{\vbox{\hrule height .#2pt
        \hbox{\vrule width.#2pt height#1pt \kern#1pt
                \vrule width.#2pt}
        \hrule height.#2pt}}}
\def\square{\mathord{\dalemb{6.8}{7}\hbox{\hskip1pt}}}
\newcommand{\ho}[1]{$\, ^{#1}$}
\newcommand{\hoch}[1]{$\, ^{#1}$}
\newcommand{\bea}{\begin{eqnarray}}
\newcommand{\eea}{\end{eqnarray}}
\newcommand{\ra}{\rightarrow}
\newcommand{\lra}{\longrightarrow}
\newcommand{\Lra}{\Leftrightarrow}
\newcommand{\ap}{\alpha^\prime}
\newcommand{\bp}{\tilde \beta^\prime}
\newcommand{\tr}{{\rm tr} }
\newcommand{\Tr}{{\rm Tr} }
\def\0{{\sst{(0)}}}
\def\1{{\sst{(1)}}}
\def\2{{\sst{(2)}}}
\def\3{{\sst{(3)}}}
\def\4{{\sst{(4)}}}
\def\5{{\sst{(5)}}}
\def\6{{\sst{(6)}}}
\def\7{{\sst{(7)}}}
\def\8{{\sst{(8)}}}
\def\n{{\sst{(n)}}}
\def\cA{{{\cal A}}}
\def\cF{{{\cal F}}}
\def\tV{\widetilde V}
\def\tW{\widetilde W}
\def\tH{\widetilde H}
\def\tE{\widetilde E}
\def\tF{\widetilde F}
\def\tA{\widetilde A}
\def\im{{{\rm i}}}
\def\tY{{{\wtd Y}}}
\def\ep{{\epsilon}}
\def\vep{{\varepsilon}}
\def\R{\rlap{\rm I}\mkern3mu{\rm R}}
\def\bD{{{\bar D}}}

\def\R{\rlap{\rm I}\mkern3mu{\rm R}}
\def\bD{{{\bar D}}}
\def\R{{{\Bbb R}}}
\def\C{{{\Bbb C}}}
\def\H{{{\Bbb H}}}
\def\CP{{{\Bbb C}{\Bbb P}}}
\def\RP{{{\Bbb R}{\Bbb P}}}
\def\Z{{{\Bbb Z}}}
\def\bA{{{\Bbb A}}}
\def\bB{{{\Bbb B}}}
\def\bC{{{\Bbb C}}}
\def\bR{{{\Bbb R}}}
\def\bD{{{\Bbb D}}}
\def\bE{{{\Bbb E}}}
\def\bZ{{{\Bbb Z}}}
\def\cD{{{\cal D}}}
\def\Re{{{\frak{Re}}}}
\def\Im{{{\frak{Im}}}}
\def\cosec{{\,\hbox{cosec}\,}}
\def\Gm{{\Gamma_{\!\! -}}}
\def\Gp{{\Gamma_{\!\! +}}}

\def\stan{{standard }}
\def\nonstan{{supernumerary }}

\def\cosech{{\hbox{cosech}}}

\def\etcyc{{\hbox{and cyclic}}}
\def\btheta{{\bar\theta}}

\newcommand{\tamphys}{\it Center for Theoretical Physics,
Texas A\&M University, College Station, TX 77843, USA}
\newcommand{\umich}{\it Michigan Center for Theoretical Physics,
University of Michigan\\ Ann Arbor, MI 48109, USA}
\newcommand{\upenn}{\it Department of Physics and Astronomy,\\
University of Pennsylvania, Philadelphia,  PA 19104, USA}
\newcommand{\SISSA}{\it  SISSA-ISAS and INFN, Sezione di Trieste\\
Via Beirut 2-4, I-34013, Trieste, Italy}

\newcommand{\mitchell}{\it George P. \& Cynthia W. 
Mitchell Institute for Fundamental Physics,\\
Texas A\&M University, College Station, TX 77843-4242, USA}

\newcommand{\newton}{\it Isaac Newton Institute for Mathematical Sciences,\\
0 Clarkson Road,  University of Cambridge,
Cambridge CB3 0EH, UK}

\newcommand{\ihp}{\it Institut Henri Poincar\'e\\
  11 rue Pierre et Marie Curie, F 75231 Paris Cedex 05}

\newcommand{\damtp}{\it DAMTP, Centre for Mathematical Sciences,
 Cambridge University,\\  Wilberforce Road, Cambridge CB3 OWA, UK}

\newcommand{\itp}{\it Institute for Theoretical Physics, University of
California\\ Santa Barbara, CA 93106, USA}

\newcommand{\gursey}{\it Feza Gursey Institute, Cengelkoy, 81220, 
Istanbul, Turkey}

\newcommand{\auth}{
G.W. Gibbons\hoch{\star \flat} and C.N. Pope\hoch{\ddagger \flat}}

\begin{document}
\begin{flushright}
\hfill {
DAMTP-2003-62\ \ \
MIFP-03-14\ \ \
}\\
\hfill{
\bf hep-th/0307052}
\end{flushright} 

\begin{center}  

{\large {\bf Consistent $S^2$ Pauli Reduction of Six-dimensional
Chiral Gauged Einstein-Maxwell Supergravity}}   

\vspace{12pt}

\auth

\vspace{7pt}
 {\hoch{\star}\damtp}

\vspace{7pt}
{\hoch{\ddagger}\mitchell}

\vspace{7pt}
{\hoch{\flat}\gursey}

\vspace{14pt}

\underline{ABSTRACT}
\end{center}  

    Six-dimensional $N=(1,0)$ Einstein-Maxwell gauged supergravity is
known to admit a (Minkowski)$_4\times S^2$ vacuum solution with
four-dimensional $N=1$ supersymmetry. The massless sector
comprises a supergravity multiplet, an $SU(2)$ Yang-Mills vector
multiplet, and a scalar multiplet.  In this paper it is shown that,
remarkably, the six-dimensional theory admits a fully consistent
dimensional reduction on the 2-sphere, implying that all solutions of
the four-dimensional $N=1$ supergravity can be lifted back to
solutions in six dimensions.  This provides a striking realisation of
the idea, first proposed by Pauli, of obtaining a theory that includes
Yang-Mills fields by dimensional reduction on a coset space.  We address
the cosmological constant problem within this model, and find that 
if the Kaluza-Klein mass scale is taken to be $10^{-3}$ eV (as has recently
been suggested) then four-dimensional gauge-coupling constants for bulk
fields must be of the order of $10^{-31}$.  We also suggest
a link between a modification of the model with 3-branes, and a 
five-dimensional model based on an $S^1/Z_2$ orbifold.

{\vfill\leftline{}\vfill
\vskip 10pt \footnoterule
{\footnotesize \hoch{\ddagger}
Research supported in part by DOE grant DE-FG03-95ER40917
\vskip -12pt} \vskip 14pt
}

\pagebreak
\setcounter{page}{1}

\newpage

\section{Introduction}

     Dimensional reductions can be divided into two types.  The first
includes the original $S^1$ reduction of Kaluza
\cite{kaluza}, and the group-manifold reductions pioneered by DeWitt
\cite{dewitt}.  All these are characterised by the fact
that the reduction ans\"atze for the metric and the other
higher-dimensional fields are invariant under a transitively-acting
group of isometries in the internal space; $U(1)$ in the case of the
Kaluza $S^1$ reduction, and $G_L$ (the left action of $G$) in the case
of a DeWitt reduction on the group manifold $G$.  The group invariance
of the ansatz ensures that the reduction will necessarily be {\it
consistent}, in the sense that all solutions of the reduced
lower-dimensional equations of motion will correspond to solutions of
the higher-dimensional equations of motion.  The transitivity of the
group action implies that there will be just a finite number of fields
in the lower-dimensional theory.  These include a $U(1)$ gauge boson
in the Kaluza reduction, and Yang-Mills fields with gauge group $G_R$
in the DeWitt reduction.

   A second type of dimensional reduction was first proposed by Pauli
\cite{strau,oraf}, ten years before DeWitt's non-abelian generalisation of
Kaluza's circle reduction.  Pauli's specific example was a reduction
on $S^2$, and more generally one can consider a reduction on any coset
space $G/H$.  In a small-fluctuation analysis there will always be
Yang-Mills fields whose gauge group is the isometry of the coset space
(which we can generally take to be $G_R$).  If these could be present
also in a full non-linear reduction, it would provide a more
economical way of obtaining Yang-Mills fields, since the number of
extra dimensions needed to obtain the gauge group $G$ from a coset
$G/H$ can be much smaller than the number needed for a DeWitt
reduction on $G$.  However, in a Pauli reduction on $G/H$ it is clear
that an ansatz that retains the Yang-Mills $G_R$ gauge fields cannot
be invariant under any transitively-acting group of isometries, and so
there is no straightforward group-theoretic reason why such a
reduction should be consistent.  In fact in general such coset
reductions are guaranteed to be {\it inconsistent}.  (Pauli never 
exhibited a concrete example with a consistent reduction.)  What is quite
remarkable, however, is that there do exist exceptional cases where
Pauli reductions can be consistent.  This gives a considerable impetus
to the search for consistent Pauli reductions.

   One of the first examples of a consistent Pauli reduction was the
reduction of eleven-dimensional supergravity on $S^7$, to give
$SO(8)$-gauged $N=8$ supergravity in four dimensions.  In fact this
example is of immense complexity, and although a proof of its
consistency is presented in \cite{dewnic}, no complete and fully
explicit reduction ansatz has been given.  A simpler, but still highly
non-trivial, example is the $S^4$ reduction of eleven-dimensional
supergravity to give gauged $SO(5)$ supergravity in seven dimensions
\cite{nasvamvan}.  A variety of other explicit examples have also been
encountered, including a consistent reduction of the
$SL(2,\R)$-singlet sector of type IIB supergravity on $S^5$, giving an
$SO(6)$ gauge theory in five dimensions \cite{s5red}, and a consistent
Pauli reduction of the $D$-dimensional bosonic string on $S^3$ or $S^{D-3}$
\cite{spherered}.  

    A further class of consistent Pauli reductions found in
\cite{spherered} consisted of $S^2$ reductions of an
Einstein-Maxwell-dilaton system in any dimension $D$, with a specific
value of the dilaton-coupling constant $a$ in the Maxwell kinetic term 
$-\ft14 e^{a\phi}\, F^2$.  This value is in fact precisely the one
that allows the Einstein-Maxwell-dilaton theory to be itself derived
as a Kaluza reduction of pure Einstein gravity in $(D+1)$ dimensions.
This fact was exploited in \cite{cvgilupo} where it was shown that the
consistent Pauli reduction on $S^2$ could be derived by starting from
the necessarily consistent DeWitt reduction of $(D+1)$-dimensional
Einstein gravity on $S^3$, and reinterpreting it as first a Kaluza
reduction to give the Einstein-Maxwell-dilaton theory in $D$
dimensions, followed by the Pauli reduction on $S^2$.  An immediate
generalisation shows that {\it any} theory obtainable from a Kaluza $S^1$
reduction of some yet higher-dimensional theory can then be
consistently Pauli reduced on $G/U(1)$ for any $G$ \cite{cvgilupo}.  A
further generalisation expressing a DeWitt reduction on $G$ as a
DeWitt reduction on $H$ followed by a Pauli reduction on $G/H$ can
also be given \cite{cvgilupo}.  Thus in these types of Pauli
reductions, one can after all find a group-theoretic understanding for
their consistency \cite{cvgilupo}.

    Not every consistent Pauli reduction admits such an interpretation
in terms of a DeWitt reduction from a yet higher dimension, however.  For
example, there is no explanation for the consistent $S^7$ reduction of
eleven-dimensional supergravity in terms of an $SO(7)$ DeWitt
reduction of some 32-dimensional theory.  In fact, none of the other
consistent reductions on $S^n$ with $n\ge3$ described above seem to 
admit any explanation in terms of DeWitt reductions.

    In this paper, we find a new and very remarkable consistent Pauli
reduction with an $S^2$ internal space.  Specifically, we consider the
six-dimensional chiral gauged supergravity studied in \cite{salsez},
which was shown to admit a supersymmetric (Minkowski)$_4\times S^2$
vacuum with $N=1$ four-dimensional supersymmetry.  The
six-dimensional theory is a particular case of a more general class of
supergravities constructed in \cite{nishsez}.  We find that
because of intricate ``conspiracies'' between the structure of the
six-dimensional theory and the properties of the 2-sphere, there
exists a consistent reduction that yields the $N=1$ supergravity coupled
to an $SU(2)$ Yang-Mills multiplet and a scalar multiplet in
four dimensions.  This reduction is all the more remarkable because
the six-dimensional gauged supergravity itself seems to have no
seven-dimensional origin via a Kaluza $S^1$ reduction.  Thus it
provides the first example of a consistent $S^2$ reduction that has no
apparent underlying group-theoretic explanation.  As such, it can
perhaps be considered to provide the most striking realisation of Pauli's
original idea.

   We begin in section 2 by presenting our $S^2$ reduction of the
bosonic sector of the chiral six-dimensional gauged Einstein-Maxwell
supergravity, indicating how the verification of the consistency of
the reduction proceeds.  In section 3 we consider the fermionic sector
of the theory, showing how one can use the six-dimensional
supersymmetry transformation rules in order to derive the fermionic
reduction ansatz.  In section 4, we consider the lift to six
dimensions of black holes in the reduced four-dimensional theory. In
section 5 we examine some of the implications of our results for
four-dimensional physics, and we make a comparison with previous
results in the literature \cite{quevedo,quevedo2}.  The model requires
dimensionless Yang-Mills coupling constants for bulk gauge fields of
the order of $10^{-31}$ if, as suggested in \cite{quevedo2}, the
Kaluza-Klein mass scale is taken to be of order $10^{-3}$ eV.  We also
find that 3-brane modifications of the model may have bulk gauge
fields with Yang-Mills coupling constants of order unity, but this
requires fine-tuning, which can be achieved with needle-shaped
internal spaces (rather than rugby balls), which are effectively
one-dimensional $S^1/Z_2$ orbifolds.  Section 6 contains discussion
and conclusions.  In an appendix, we give some details of the
curvature calculations for the class of metric reduction ansatz that
we use in this paper.

\section{Pauli $S^2$ Reduction of the Bosonic Sector}

     The bosonic sector of the six-dimensional $N=(1,0)$ gauged
Einstein-Maxwell supergravity is described by the Lagrangian \cite{salsez}
\be
{\cal L} = \hat R\, {\hat *\oneone} - \ft14 {\hat*d\hat\phi}
\wedge d\hat \phi  - \ft12 e^{\hat\phi}\,
{\hat *\hat H_\3}\wedge \hat H_\3 - \ft12 e^{\ft12\hat \phi}\, 
{\hat *\hat F_\2}\wedge \hat F_\2
 - 8g^2 \, e^{-\ft12\hat \phi}\, {\hat *\oneone}\,,\label{d6lag}
\ee
where $\hat F_\2= d\hat A_\1$, $\hat H_\3 = d\hat B_\2 + 
\ft12 \hat F_\2\wedge \hat A_\1$, and we place hats on all six-dimensional
fields.  (We use conventions where ${\hat *\hat\omega}\wedge \hat\omega
= (1/p!)\, \hat\omega^{M_1\cdots M_p}\, \hat\omega_{M_1\cdots M_p}\, 
{\hat *\oneone}$ for any $p$-form $\hat\omega$.)  Here $g$ is
the gauge-coupling constant, and the fermions all carry charge $g$ in 
their minimal coupling to the $U(1)$ gauge field $\hat A$.  
The bosonic equations of motion following from (\ref{d6lag}) are
\bea
\hat R_{MN} &=& \ft14 \del_M\hat \phi\, \del_N\hat \phi + 
\ft12 e^{\ft12\hat \phi}\,
(\hat F^2_{MN} - \ft18 \hat F^2\,\hat g_{MN}) + \ft14 e^{\hat \phi}\,
(\hat H^2_{MN} - \ft16 \hat H^2\, \hat g_{MN})\nn\\
&& + 2g^2\,  e^{-\ft12\hat \phi}\, \hat g_{MN}\,,\nn\\
\hat{\square}\, \hat \phi &=& \ft14 e^{\ft12\hat \phi}\, \hat F^2 + 
\ft16 e^{\hat \phi}\, \hat H^2 -
     8g^2\, e^{-\ft12\hat \phi}\,,\nn\\
d(e^{\ft12\hat \phi}\,{\hat *\hat F_\2}) &=& e^{\hat \phi}\, 
{\hat *\hat H_\3}\wedge \hat F_\2\,,\label{eoms}\\
d(e^{\hat \phi}\, {\hat *\hat H_\3}) &=& 0\,.\nn
\eea
Note that the dimensionful coupling constant $g$ can be rescaled at
will by adding a constant to $\hat\phi$, together with compensating
rescalings of the other fields. Thus it is really the quantity $g\,
e^{-\ft14\hat\phi_0}$, where $\hat\phi_0$ is the expectation value of
$\hat\phi$, that has physical significance in the theory.  We could,
conversely, without loss of generality set $\hat \phi_0=0$.

    It has long been known that this theory admits a solution of the
form (Minkowski)$_4\times S^2$, and furthermore, that this solution
has $N=1$ supersymmetry in the four-dimensional spacetime
\cite{salsez}.  The spectrum of four-dimensional massless fields has
been discussed at the linearised level in \cite{salsez,quevedo}, and
according to \cite{quevedo} it comprises the $N=1$ supergravity
multiplet, a Yang-Mills $SU(2)$ triplet of vector multiplets, and a
singlet scalar multiplet.  The $SU(2)$ Yang-Mills fields have a
natural origin in the isometry group of the compactifying 2-sphere.

   In the reduction of a generic theory on a coset space such as $S^2=
SO(3)/SO(2)$, the massless fields that one finds in a linearised
analysis of the harmonic expansion will not decouple from the massive
fields associated with the higher harmonics.  In other words, one
finds that when the full nonlinear structure of the theory is taken
into account, the field equations for the massive fields will have
source terms built purely from the massless fields.  These sources
prevent one from consistently setting the massive fields to zero.  In
the present case, however, it turns out that some very remarkable
conspiracies imply that there is an exact decoupling of the massive
fields, allowing us to find a consistent reduction on $S^2$ that
yields well-defined four-dimensional equations of motion for the
massless sector of the full dimensionally-reduced theory.  In
particular, we obtain the $SU(2)$ Yang-Mills fields associated with
the isometry group of the round 2-sphere, despite the fact that the
reduction ansatz will (necessarily) not be invariant under the $SU(2)$
action of the round sphere.

   We arrived at the correct reduction ansatz by a process of 
trial and error, and adjusting of certain parameters.  We shall now simply
present the result, and then indicate how one establishes that it does indeed 
work. The six-dimensional metric will be written as
\be
d\hat s^2 = e^{\ft12\phi}\, ds_4^2 + e^{-\ft12\phi}\, 
g_{mn}\, (dy^m + 2g\, A^i\, K_i^m)(dy^n + 2g\, A^j\, K_j^n)\,,\label{metans}
\ee
where $g_{mn}$ is the metric on the round $S^2$, normalised to $R_{mn}=
8g^2\, g_{mn}$, where $g$ is the gauge coupling constant in
(\ref{d6lag}).  The quantities $K_i=K_i^m\, \del/\del y^m$ are the
three Killing vectors on the round 2-sphere.  The four-dimensional
metric $ds_4^2= g_{\mu\nu}\, dx^\mu\, dx^\nu$, the Yang-Mills gauge
potentials $A^i=A^i_\mu\, dx^\mu$, and the scalar field $\phi$ all
depend on the four-dimensional coordinates $x^\mu$ only, and are
independent of the coordinates $y^m$ of the 2-sphere.  It will prove
convenient to introduce an orthonormal basis $\hat e^A$ for the
six-dimensional metric, which we do by defining
\be
\hat e^\a = e^{\ft14\phi}\, e^\a\,,\qquad 
\hat e^a = e^{-\ft14\phi}\, (e^a + 2g\, A^i\, K_i^a)\,,\label{vielans}
\ee
where $e^\a$ is an orthonormal basis for $ds_4^2$, and $e^a$ is an
orthonormal basis for the metric $g_{mn}\, dy^m\, dy^n$ on the round 
2-sphere.  The quantities $K_i^a$ are the orthonormal frame components of the
Killing vectors on $S^2$: $K_i^a = K_i^m\, e^a_m$.

   We find that the appropriate ansatz for the other six-dimensional
fields is as follows:
\bea
\hat F_\2 &=& 2 g\, e^{\ft12\phi}\,\ep_{ab}\, \hat e^a\wedge \hat e^b
- \mu_i\, F^i\,,\nn\\   
\hat H_\3 &=& H_\3 
- 2 g\, F^i\wedge K_i^a\, (e^a +2g\, A^j\, K_j^a)\,,\label{fhphi}\\
\hat \phi &=& -\phi\,,\nn
\eea
where the Yang-Mills field strengths are defined by $F^i = dA^i + g\,
\ep_{ijk}\, A^j\wedge A^k$.
Here, the three scalars $\mu_i$ are the $SU(2)$ triplet of lowest non-trivial
harmonics on $S^2$.  It will prove convenient for some purposes to have
an explicit representation for the metric of the 2-sphere.  Bearing in
mind that the round $S^2$ in (\ref{metans}) is normalised so that
$R_{mn}= 8g^2\, g_{mn}$, it follows that we may write it as
\be
g_{mn}\, dy^m\, dy^n = \fft{1}{8g^2}\, 
(d\theta^2 + \sin^2\theta\, d\varphi^2)\,.
\ee
In terms of these coordinates, we can write the scalar harmonics $\mu_i$
as
\be
\mu_1 = \sin\theta\, \cos\varphi\,,\qquad
\mu_2 = \sin\theta\, \sin\varphi\,,\qquad
\mu_3 = \cos\theta\,.
\ee
One can easily construct the Killing vectors $K_i^m$ from these, as
\be
K_i^m = \fft1{8g^2}\, \ep^{mn}\, \del_n\, \mu_i\,,\label{kvcon}
\ee
giving
\be
K_1 = -\sin\varphi\, \fft{\del}{\del\theta} - \cot\theta\, \cos\varphi\, 
\fft{\del}{\del\varphi}\,,\quad
K_2 = \cos\varphi\, \fft{\del}{\del\theta} - \cot\theta\, \sin\varphi\, 
\fft{\del}{\del\varphi}\,,\quad
K_3= \fft{\del}{\del\varphi}\,.
\ee
They satisfy the algebra $[K_i,K_j]= -\ep_{ijk}\,K_k$.
Some useful {\it lemmata} are
\bea
&&K_i = \fft{1}{8g^2}\, \ep_{ijk}\, \mu_j\, d\mu_k\,,\qquad 
dK_i = 8g^2\, \ep_{ijk}\, 
K_j\wedge K_k = \fft1{4g^2}\, \mu_i\, \Omega_\2\,,\label{lemmata}\\
&&g^{mn}\, \del_m \mu_i\, \del_n\mu_j = 8g^2\, (\delta_{ij} - \mu_i\,
\mu_j)\,,\qquad 
\ep^{mn}\, \del_m\mu_i\, \del_n\mu_j = 8g^2\, \ep_{ijk}\, \mu_k\,,
\nn
\eea
where $\Omega_\2\equiv \sin\theta\, d\theta\wedge d\varphi$ is the volume form
of the unit $S^2$.

   In performing the substitution of the ansatz into the
six-dimensional field equations, it is useful to note that the
six-dimensional Hodge duals of $\hat F_\2$ and $\hat H_\3$ are given
by
\bea
{\hat * \hat F_\2} &=& 4g\, e^{\ft32\phi} \, {*\oneone} - \ft12\, \mu_i\,
\ep_{ab}\, {*F^i}\wedge \hat e^a\wedge \hat e^b\,,\nn\\
{\hat * \hat H_\3} &=& \ft12 e^{-\ft12\phi}\, \ep_{ab} {* H_\3}\wedge 
\hat e^a\wedge \hat e^b  + \fft1{4g}\, {*F^i}\wedge D\mu_i\,,
\label{duals}
\eea
where an unhatted $*$ denotes the Hodge dual in the four-dimensional 
metric $ds_4^2$, and
\be
D\mu_i \equiv d\mu_i + 2 g\, \ep_{ijk}\, A^j\, \mu_k\,.
\ee

   The results from substituting the ansatz into the six-dimensional 
Bianchi identities and field equations are as follows.  First, one can
see from (\ref{fhphi}) that $\hat F_\2$ can be rewritten as
$\hat F_\2= (2g)^{-1}\, \Omega_\2 - d(\mu_i\, A^i)$, which shows that
$d\hat F_\2=0$ is satisfied identically.  After some algebra one finds
that the Bianchi identity $d\hat H_\3=\ft12\hat F_\2\wedge \hat F_2$
implies the four-dimensional equation
\be
dH_\3 = \ft12 F^i\wedge F^i\,.
\ee
The six-dimensional equation $d(e^{\hat\phi}\,{\hat *\hat H_\3})=0$ implies
the two four-dimensional equations
\bea
D(e^{-\phi}\, {*F^i}) &=& e^{-2\phi}\, {*H_\3}\wedge F^i\,,\nn\\
d(e^{-2\phi}\, {* H_\3}) &=& 0\,,\label{fheqs}
\eea
while the six-dimensional equation $d(e^{\ft12\hat\phi}\, {\hat *\hat
F^i}) = e^{\hat\phi}\, {\hat * \hat H_\3}\wedge \hat F_2$ gives again
the first of the equations in (\ref{fheqs}).  The six-dimensional 
dilaton equation of motion yields the four-dimensional equation
\be
d{*d\phi} = \ft12 e^{-\phi}\, {*F^i}\wedge F^i + e^{-2\phi}\, 
{*H_\3}\wedge H_\3\,.
\ee

   Finally, using the expressions for the Ricci tensor given in
(\ref{ricci2}), we find that the six-dimensional Einstein equations
involving $\hat R_{\a\beta}$, $\hat R_{\a b}$ and $\hat R_{ab}$ 
respectively yield the four-dimensional equations
\bea
R_{\a\beta} &=& \ft12\nabla_\a\phi\, \nabla_\beta\phi +
\ft14\square\phi\, \eta_{\a\beta} + \ft12 e^{-\phi} \, 
(F^i_{\a\gamma}\, F^i_\beta{}^\gamma - \ft18 F^i_{\gamma\delta}\, 
   F^{i\, \gamma\delta}\, \eta_{\a\beta}) \nn\\
&& + \ft14 e^{-2\phi}\, (H^2_{\a\beta} - \ft16 H^2\,
\eta_{\a\beta})\,,\nn\\
D^\beta(e^{-\phi}\, F^i_{\a\beta}) &=& -\ft12 e^{-2\phi}\,
H_\a{}^{\beta\gamma}\, F^i_{\beta\gamma}\,,\label{einstcon}\\
\square \phi &=& -\ft14 e^{-\phi}\, F^i_{\a\beta}\, F^{i\, \a\beta} 
-\ft16 e^{-2\phi}\, H^2\,.\nn
\eea

   We find that the system of four-dimensional equations that we have
obtained by this 2-sphere reduction can be derived from an action
principle, with the Lagrangian given by
\be
{\cal L} = R\, {*\oneone} - \ft12 {*d\phi}\wedge d\phi - \ft12
e^{-\phi}\, {*F^i}\wedge F^i -\ft12 e^{-2\phi}\, {*H_\3}\wedge H_\3\,,
\label{4lag}
\ee
where $F^i = dA^i + g\, \ep_{ijk}\, A^j\wedge A^k$ and $H_\3 = dB_\2 
+ \ft12 \omega_\3$, where $d\omega_\3 = F^i\wedge F^i$.     

     It should be emphasised that highly non-trivial cancellations
take place when one substitutes the ansatz into the higher-dimensional 
field equations.  In particular, it is only because of the specific 
details of how the Yang-Mills fields enter not only in the metric
ansatz, but also in the ans\"atze for $\hat F_\2$ and $\hat H_\3$ that
one finds that all the dependence on the coordinates $y^m$ on $S^2$
eventually cancels out, leading to well-defined four-dimensional
equations of motion.   This point can be illustrated by examining some
of the calculations for the reduction of the Einstein equations in
more detail.  For the components $\hat R_{\a\beta}$ in (\ref{eoms}), 
substitution of the ans\"atze gives
\bea
R_{\a\beta}&=& \ft12\nabla_\a\phi\, \nabla_\beta\phi +\ft14
\square\phi\, \eta_{\a\beta} + \ft12e^{-\phi} \, F^i_{\a\gamma}\, 
  F^j_\beta{}^\gamma \, (\mu_i\, \mu_j + 4 g^2\, K^a_i\, K_{j\, a} 
 + 4 g^2\, K^a_i\, K_{j\, a}) \nn\\
&& -\ft1{16} e^{-\phi} \, F^i_{\gamma\delta}\, 
  F^{j\, \gamma\delta} \,( \mu_i\, \mu_j + 8 g^2\, K^a_i\, K_{j\, a})
+ \ft14 e^{-2\phi}\, (H^2_{\a\beta} - \ft16 H^2\, \eta_{\a\beta})\,,
\label{albet}
\eea
where we have kept distinct the contributions in the $F^i\, F^j$ terms
coming from the different sources. In the $F^i_{\a\gamma}\,
F^j_\beta{}^\gamma$ terms, the $\mu_i\, \mu_j$ contribution comes from
the ansatz for $\hat F_\2$, while the two $4 g^2\, K^a_i\, K_{j\, a}$
contributions come from the ans\"atze for $\hat H_\3$ and the metric.
In the $F^i_{\gamma\delta}\, F^{j\, \gamma\delta}$ the two
contributions come from the ans\"atze for $\hat F_\2$ and $\hat H_\3$
respectively.  Using (\ref{kvcon}) and (\ref{lemmata}), one finds that
$\mu_i\, \mu_j + 8g^2\, K^a_i\, K_{j\, a} = \delta_{ij}$, and hence
all the $y$-dependence cancels in (\ref{albet}), yielding the first
equation in (\ref{einstcon}).  The components $\hat R_{ab}$ of the
Einstein equations in (\ref{eoms}) provide another
cancellation, with all the $y$-dependence associated with the
$F^i_{\gamma\delta}\, F^{j\, \gamma\delta}$ terms again conspiring to
vanish.  A further remarkable feature evident in these components is that
the consistent reduction is achieved with only the single ``breathing mode
scalar'' $\phi$ in the reduction ansatz.  Usually, the scalars
parameterising deformations of the internal space would be needed in
any consistent reduction, since the Yang-Mills fields would act as
sources for them.  In our present example, these source terms turn out
to cancel, and so the normally-expected 5 modulus scalars are consistently set
to zero.

   The four-dimensional theory that we have arrived at can be cast
into a more conventional form by performing a dualisation of $H_\3$ to
an axionic scalar.  We do this by the standard procedure of
introducing a Lagrange multiplier $\sigma$ and adding the term
$-\sigma\, (dH_\3 - \ft12 F^i\wedge F^i)$ to (\ref{4lag}).  This
imposes the Bianchi identity for $H_\3$.  Varying instead with respect
to $H_\3$, we find that
\be
H_\3 = e^{2\phi}\, {*d\sigma}\,.\label{dual}
\ee
Substituting this into the modified Lagrangian gives the dualised version
\be
{\cal L} = R\, {*\oneone} - \ft12 {*d\phi}\wedge d\phi 
-\ft12 e^{2\phi}\, {*d\sigma}\wedge d\sigma - \ft12
e^{-\phi}\, {*F^i}\wedge F^i + \ft12 \sigma\, F^i\wedge F^i\,.
\label{4lag2}
\ee

\section{The Fermionic Sector}\label{fermsec}

   The easiest way to determine the correct reduction ansatz in the
fermionic sector is by looking at the higher-dimensional supersymmetry
transformation rules.  For the fermions, we have
\bea
\delta\hat\psi_M & =&
\hat D_M\, \hat\ep  + \ft1{48} e^{\ft12\hat\phi}\,
\hat H_{NPQ}\,\hat\Gamma^{NPQ}\, \hat\Gamma_M\, \hat\ep\,,\nn\\
\delta\hat\chi &=&
-\ft14[\del_M\hat\phi\, \hat\Gamma^M - \ft16 e^{\ft12\hat\phi}\, 
\hat H_{MNP}\, \hat\Gamma^{MNP}]\hat\ep\,,\label{susytrans}\\
\delta \hat\lambda &=&
\ft1{4\sqrt{2}}[e^{\ft14\hat\phi}\, \hat F_{MN}\, \hat\Gamma^{MN} - 8\im\,
g\, e^{-\ft14\hat\phi}]\hat\ep\,,\nn
\eea
where $\hat D_M$ is the gauge-covariant derivative, $\hat D_M\hat\ep
\equiv (\hat \nabla_M - \im\, g\, \hat A_M)\hat \ep$.  The transformation
rules for the bosons are
\bea
\delta \hat e^A_M &=& -\ft14 \hat{\bar\ep}\, \hat\Gamma^A\, \hat\psi_M 
  + \ft14 \hat{\bar\psi}_M\, \hat\Gamma^A\, \hat\ep \,,\nn\\
\delta \hat\phi &=& \ft12\hat{\bar\ep}\, \hat\chi + 
\ft12\hat{\bar\chi}\, \hat\ep
\,,\nn\\
\delta \hat A_M &=& \ft1{2\sqrt2}\, e^{-\ft12\hat\phi}\, (
\hat{\bar\ep}\, \hat\Gamma_M\, \hat\lambda - \hat{\bar\lambda}\, 
\hat\Gamma_M\, \hat\ep)\,,\label{bossusy}\\
\delta\hat B_{MN} &=& \hat A_{[M}\, \delta\hat A_{N]} +\ft12 
e^{-\hat\phi}\, (\hat{\bar\ep}\, \hat\Gamma_{[M}\, \hat\psi_{N]}
  + \hat{\bar\psi}_{[M} \, \hat\Gamma_{N]}\, \hat\ep +
\hat{\bar\ep}\, \hat\Gamma_{MN}\, \hat\chi - \hat{\bar\chi}\, 
\hat\Gamma_{MN}\, \hat\ep)\,.\nn
\eea

   Since the fermion kinetic terms in the six-dimensional Lagrangian
are of the form
\be
\hat e^{-1}\, 
{\cal L} = \hat{\bar\psi}_M\, \hat\Gamma^{MNP}\, \hat D_N\, \hat \psi_P + 
      \hat{\bar\chi}\, \hat\Gamma^M\, \hat D_M\, \hat\chi +
      \hat{\bar\lambda}\, \hat\Gamma^M\, \hat D_M\, \hat\lambda\,,
\ee
it follows that in order to obtain a four-dimensional theory with
canonical fermion kinetic terms having no dilaton exponential
scalings, the $\hat\chi$ and $\hat\lambda$ fields and the tangent-frame 
gravitino components $\hat\psi_A$ should all receive scalings by
a factor of $e^{-\ft18\phi}$ in their dimensional reductions. It then 
follows from scaling arguments that in order to obtain a canonical
four-dimensional gravitino transformation rule $\delta\psi_\mu 
= \nabla_\mu\, \ep+\cdots$, with no dilaton exponential scaling in 
the derivative term, the six-dimensional supersymmetry parameter 
$\hat\ep$ should receive a scaling by $e^{\ft18\phi}$ in its
dimensional reduction.  

   We now introduce the chiral gauge-covariantly constant 2-component spinor
$\eta$ on $S^2$, which satisfies the equation
\be
(\nabla_a - \im\, A_a^{\rm mono})\, \eta=0\,,
\ee
where $A_a^{\rm mono}$ is a potential for the Dirac monopole on $S^2$.
The six-dimensional Dirac matrices $\hat\Gamma_A$ will be decomposed
as
\be
\hat\Gamma_\a = \gamma_\a \otimes \sigma_3\,,\qquad 
\hat\Gamma_a = \oneone\otimes \sigma_a\,,
\ee
where $1\le a\le 2$.  The six-dimensional supersymmetry parameter is
then decomposed as
\be
\hat\ep = e^{\ft18\phi}\, \ep\otimes \eta\,.
\ee

   Substituting first into the transformation rule for $\hat\lambda$
in (\ref{susytrans}), and using (\ref{fhphi}), we find that
\be
\delta\hat\lambda = \sqrt2\, \im\, g\, e^{\ft38\phi}\, 
\ep\otimes (\sigma_3-1)\, \eta - \ft1{4\sqrt2}\, e^{-\ft58\phi}\, 
\mu_i\, F^i_{\a\beta}\, \gamma^{\a\beta}\ep\otimes \eta\,.
\ee
Thus we deduce that the chirality of $\eta$ is
\be
\sigma_3\, \eta = +\eta\,,
\ee
and that the dimensional reduction of $\hat\lambda$ should be given by
\be
\hat\lambda = e^{-\ft18\phi}\, \mu_i\, \lambda^i\otimes \eta\,.
\ee
We therefore obtain a purely four-dimensional expression for
$\delta\lambda^i$, with no dependence on the coordinates of $S^2$, 
namely
\be
\delta\lambda^i = -\ft1{4\sqrt2}\, e^{-\ft12\phi}\,  
F^i_{\a\beta}\, \gamma^{\a\beta}\ep\,.
\ee
The triplet of spin $\ft12$ fields $\lambda^i$ form the $N=1$
superpartners of the $SU(2)$ Yang-Mills fields $A_\mu^i$.

   Proceeding in the same vein,  we can determine the appropriate
ans\"atze for the dimensional reduction of the remaining fermionic
fields, the guiding principle being that one should thereby obtain 
consistent purely four-dimensional transformation rules, with no
dependence on the coordinates of the internal 2-sphere.  The summary
of our results is the following.  The reduction ans\"atze are
\bea
\hat\lambda &=& e^{-\ft18\phi}\, \mu_i\, \lambda^i\otimes \eta\,,\nn\\
\hat\chi &=& e^{-\ft18\phi}\, 
[\chi\otimes \eta + \ft{\sqrt2}3\, g\, K_i^a\, \lambda^i\otimes
\sigma_a  \eta]\,,\nn\\
\hat\psi_\a &=& e^{-\ft18\phi}\, [ \psi_\a\otimes \eta +\ft1{\sqrt2}\, g\,
K_i^a\, \gamma_\a \lambda^i\otimes \sigma_a\eta]\,,\label{fermans}\\
\hat\psi_a &=& e^{-\ft18\phi}\, [-\ft12 \chi\otimes \sigma_a\eta +
\ft3{\sqrt2}\, g\, K_i^a\, \lambda^i\otimes\eta - \ft{\im}{\sqrt2}\, 
g\, \ep_{ab}\, K_i^b\, \lambda^i
\otimes \eta]\,,\nn
\eea
leading to the following four-dimensional transformation rules
\bea
\delta\lambda^i &=& -\ft1{4\sqrt2}\, e^{-\ft12\phi}\,  
F^i_{\a\beta}\, \gamma^{\a\beta}\ep\,,\nn\\
\delta\chi &=& \ft14 \del_\a\phi\, \gamma^\a\, \ep + \ft1{24} e^{-\phi}\, 
H_{\a\beta\gamma}\, \gamma^{\a\beta\gamma}\, \ep\,,\label{4dimsusy1}\\
\delta\psi_\a &=&  \nabla_\a\, \ep + \ft18 \del_\beta\phi\, \gamma_\a
\gamma^\beta\, \ep + \ft1{48} e^{-\phi}\, H_{\beta\gamma\delta}\, 
\gamma^{\beta\gamma\delta}\, \gamma_\a\ep\,.\nn
\eea
As in the reductions of the bosonic fields discussed earlier, we again
find that the way in which the dependences on the internal 2-sphere
coordinates $y^m$ eventually match up, so as to allow four-dimensional 
transformation rules to be consistently read off, involves some rather
non-trivial cancellations between terms.

   The four-dimensional transformation rules (\ref{4dimsusy1}) take a
more familiar form if we rewrite them using the axion $\sigma$ dual to
$H_\3$, given by (\ref{dual}). Also, it is advantageous to redefine
the gravitino, by setting
\be 
\psi_\a = \psi_\a' +\ft12 \gamma_\a\, \chi\,.
\ee
The four-dimensional fermionic supersymmetry transformations then become
\bea
\delta\lambda^i &=& -\ft1{4\sqrt2}\, e^{-\ft12\phi}\,  
F^i_{\a\beta}\, \gamma^{\a\beta}\ep\,,\nn\\
\delta\chi &=& \ft14 (\del_\a\phi\, -\im\, e^\phi\, \del_\a \sigma\,
\gamma_5) \gamma^\a\, \ep \,,\\
\delta\psi_\a' &=& \nabla_\a\, \ep  -\ft{\im}{4}\, 
e^\phi\, \del_\a\sigma\, \gamma_5\ep\,.\nn
\eea

   If the reduction ansatz (\ref{fermans}), together with the previous
bosonic ansatz, are substituted into the six-dimensional fermionic
equations of motion, one will again find that four-dimensional
equations of motion emerge in a consistent manner.  In this regard we
remark that the gravitino $\psi_\mu$ and the spinor $\chi$ in the 
scalar multiplet $(\phi,\sigma,\chi)$ are uncharged under the 
$SU(2)$ Yang-Mills group, while the spinors $\lambda^i$ in the vector
multiplets $(A_\mu^i,\lambda^i)$ have the expected minimal coupling,
with $D_\mu\, \lambda^i = \nabla_\mu\, \lambda^i + 2 g\, \ep_{ijk}\, 
A^j_\mu\, \lambda^k$.  

   Turning now to the supersymmetry transformation rules for the
bosons, we first find by substituting into the variation of $\hat\phi$ 
in (\ref{bossusy}) that
\be
\delta\phi = - \ft12(\bar\ep\, \chi + \bar\chi\,\ep)\,.
\ee
Next, from $\delta\hat A_M$ we find the four-dimensional transformations
\be
\delta A^i_\mu = -\ft1{2\sqrt2}\, e^{\ft12\phi}\, 
(\bar\ep\, \gamma_\mu\, \lambda^i - \bar\lambda^i\, \gamma_\mu\, \ep)\,.
\ee

    As usual in a dimensional reduction, it is necessary when
analysing the supersymmetry transformation of the vielbein to perform
also a compensating local Lorentz transformation in order to maintain
the gauge choice $\hat e^\a_m=0$ of the reduction ansatz
(\ref{vielans}), and to allow the six-dimensional supersymmetry
transformations to be interpreted in terms of transformations of the
four-dimensional fields. Thus we augment $\delta\hat e^A_M$ in
(\ref{bossusy}) to
\be
\delta\hat e^A_M = -\ft14 \hat{\bar\ep}\, \hat\Gamma^A\, \hat\psi_M 
  + \ft14 \hat{\bar\psi}_M\, \hat\Gamma^A\, \hat\ep 
 + \Lambda^A{}_B\, \hat e^B_M\,,
\ee
where $\Lambda_{AB}=-\Lambda_{BA}$, and take
\bea
\Lambda^\a{}_b &=& \ft{3}{4\sqrt2}\, g \, K_{i\, b}\, 
(\bar\ep\, \gamma^\a\, \lambda^i -\bar\lambda^i\, \gamma^\a\, \ep) 
-\ft{\im}{2\sqrt2}\, g\, \ep_{bc}\, K_i^c\,(\bar \ep\, \gamma^\a\, \lambda^i 
-\bar\lambda^i\, \gamma^\a\, \ep)\,,\nn\\
\Lambda^a{}_b &=& -\ft{\im}{8}\, \ep^a{}_b \, (\bar\ep\, \chi - 
\bar\chi\, \ep)\,.
\eea
Additionally,
it is advantageous to perform a further purely four-dimensional 
local Lorentz transformation, whose purpose is simply to neaten up 
and simplify the results:  
\be
\Lambda^\a{}_\beta = \ft18 (\bar\ep \, \gamma^\a{}_\beta \, \chi
-\bar\chi\, \gamma^\a{}_\beta\, \ep)\,.
\ee
The various four-dimensional sectors of the six-dimensional vielbein
transformation rules then reproduce previously-derived results for
$\delta\phi$ and $\delta A_\m$, and give the four-dimensional vierbein
transformation
\be
\delta e^\a_\mu = -\ft14 (\bar\ep\, \gamma^\a\, \psi_\mu' -
    \bar\psi_\mu'\, \gamma^\a\, \ep)\,.
\ee

\section{Uplifting of Four-Dimensional Black Holes}

    Since we have obtained a consistent Pauli reduction of the
six-dimensional theory, it follows that any solution in four
dimensions can be lifted back to give a solution in six dimensions.  A
case of interest is an extremal black hole in four dimensions,
carrying a magnetic charge supported by a field in the abelian $U(1)$ 
subgroup of the $SU(2)$ Yang-Mills.  Thus taking $F^i$ to be zero
except for $F^3 =F$,  in four dimensions we shall have
\bea
ds_4^2 &=& - {\cal H}^{-1}\, dt^2 + {\cal H}\, [d\rho^2 + 
\rho^2\, (d\td\theta^2 + \sin^2\td\theta\, d\td\varphi^2)]\,,\nn\\
e^\phi &=& {\cal H}\,,\qquad F = -Q\, \sin\td\theta\, d\td\theta\wedge
d\td\varphi\,,\qquad {\cal H} = 1 + \fft{Q}{\rho}\,.
\eea
Note that although this solution is extremal it is not supersymmetric
within the $N=1$ theory, since the charge is carried by a gauge field
in a matter multiplet.  (See (\ref{4dimsusy1}).)

   Lifting to six dimensions using the reduction formulae (\ref{metans})
and (\ref{fhphi}), we find that the metric is given by
\be
d\hat s_6^2 \!=\!\! - {\cal H}^{-\ft12}\, dt^2 + {\cal H}^{\ft32}\, [
d\rho^2 + \rho^2\, (d\td\theta^2 + \sin^2\td\theta\, d\td\varphi^2)] 
+ \fft1{8g^2}\, {\cal H}^{-\ft12}\, [d\theta^2 + \sin^2\theta\, 
(d\varphi + 2 g\, Q\, \cos\td\theta\, d\td\varphi)^2]\,,
\label{lifted1}
\ee
while the other six-dimensional fields are given by
\bea
\hat F_\2 &=& \fft1{2g}\, \sin\theta\, d\theta\wedge (d\varphi + 
2g\, Q\, \cos\td\theta\, d\td\varphi) + Q\, \cos\theta\,
\sin\td\theta\, d\td\theta\wedge d\td\varphi\,,\nn\\
\hat H_\3 &=& \fft{Q}{4g}\, \sin\td\theta\, \sin^2\theta\,
d\td\theta\wedge d\td\varphi\wedge d\varphi\,,\label{lifted2}\\
e^{\hat\phi} &=& {\cal H}^{-1}\,.\nn
\eea

    Since the azimuthal coordinate $\varphi$ on the Pauli reduction
2-sphere has period $2\pi$, it follows that the 1-form $(d\varphi + 2
g\, Q\, \cos\td\theta\, d\td\varphi)$ appearing in (\ref{lifted1}) and
(\ref{lifted2}) will be globally defined if 
\be
2 g\, Q = \ft12 n\,,
\ee
where $n$ is an integer.  However, the six-dimensional metric will 
still become singular on the horizon at $\rho=0$.  We can study the
near-horizon limit by taking ${\cal H} \longrightarrow Q/\rho$.
Defining a new radial coordinate by $r = 4 \rho^{\ft14}\, Q^{\ft34}$, we
find that in the near-horizon limit (\ref{lifted1}) approaches
\be
d\hat s_6^2 = dr^2 + r^2\, \Big\{ -\fft{dt^2}{16Q^2} + \fft1{8n^2}\, 
[ d\theta^2 + \sin^2\theta\, (d\varphi + \ft12 n\, \cos\td\theta\, 
d\td\varphi)^2 + \ft12 n^2\, (d\td\theta^2 + \sin^2\td\theta\,
d\td\varphi^2) ]\Big\}\,.
\ee
Thus in this limit the metric approaches a cone over $\R$ times an
$S^2$ bundle over $S^2$.  If $n$ is odd the bundle is non-trivial,
while if $n$ is even it is trivial.

\section{Four-dimensional Physics and Compactification Scale}

   Having shown in this paper that there exists an exact consistent
Pauli reduction of the six-dimensional gauged $N=(1,0)$ supergravity, it is 
of interest to examine in further detail the resulting four-dimensional 
$N=1$ theory, and its relation to the six-dimensional starting point.

  The four-dimensional theory admits a Minkowski vacuum, in which the
dilaton field $\phi$ is a constant.  As can be seen from (\ref{4lag}),
the value of this constant, say $\phi=\phi_0$, can be arbitrary.  It
is evident from the metric reduction ansatz (\ref{metans}) that the
geometric radius of the compactifying 2-sphere, as measured in the
six-dimensional metric, will then be given by
\be
{\cal R} = \fft1{2\sqrt2\, g}\, e^{-\ft14\phi_0}\,.\label{radius1}
\ee
From the form of the four-dimensional Lagrangian (\ref{4lag}), we see that
we should rescale the $SU(2)$ Yang-Mills fields $A^i$, and hence also the
gauge-coupling constant $g$, by appropriate powers of $e^{\phi_0}$ in order
to work with canonically-normalised fields.  Thus, we define 
\be
\wtd A^i = e^{-\ft12\phi_0}\, A^i\,,\qquad 
\td g = e^{\fft12\phi_0}\, g\,,\label{rescale}
\ee
in terms of which the Lagrangian expanded around the $\phi=\phi_0$ 
background will take the form
\be
{\cal L} = R\, {*\oneone} - \ft12 {*\wtd F^i}\wedge \wtd F^i + \cdots\,,
\label{action3}
\ee
where $\wtd F^i = d \wtd A^i + \td g\, \ep_{ijk}\, \wtd A^j\wedge \wtd A^k$.
Thus the radius ${\cal R}$ and the rescaled  
Yang-Mills coupling constant $\td g$ are related by
\be
{\cal R} = \fft1{2\sqrt2\, \td g}\, e^{+\ft14\phi_0}\,,\label{radius2}
\ee
which should be contrasted with the expression (\ref{radius1})
involving the gauge-coupling $g$ in the non-canonical normalisation.
Note that the coupling constant $\td g$ has dimensions of inverse
length.  When the action (\ref{action3}) is converted to standard
units, one obtains a classical Yang-Mills coupling constant $g_{\rm
YM}^{\phantom \Sigma} = \td g\, (\sqrt{4\pi G})/c$ which again is
dimensionful.  Quantum perturbation theory is an expansion in powers of
the dimensionless constant $\hbar\, g_{\rm YM}^2/(4\pi\, c) = (\td g\,
L_{\rm planck})^2$.  Thus if we wanted the $SU(2)$ gauge-coupling
constant $g_{\rm YM}$ to be of order unity, we should take 
$\td g \sim 1/L_{\rm planck}$.

    Note that one can include a Yang-Mills
sector in the original $D=6$ theory, implying additional terms 
\be
{\cal L}_{\rm YM} = -\ft12 e^{\ft12\hat\phi}\, {\hat *\hat F^I}\wedge 
\hat F^I
\ee
that are added to (\ref{d6lag}), where $\hat F^I = d\hat A^I + 
g'\, f_{JK}{}^I\, \hat A^J\wedge \hat A^K$.  These fields reduce to
four dimensions simply by setting $\hat A^I = A^I$, implying additional terms
\be
{\cal L} = -\ft12 e^{-\phi}\, {*F^I}\wedge F^I
\ee
in (\ref{4lag}).  These fields might
be taken to have the gauge group $E_6\times E_7$, with the fields of
the standard model being embedded in them.  Conversion to
canonically-normalised fields in the $\phi=\phi_0$ vacuum would again
require rescalings
\be
\wtd A^I = e^{-\ft12\phi_0}\, A^I\,,\qquad \td g' = e^{\ft12\phi_0}\, 
g'\,,\label{rescale2}
\ee
just as for the $SU(2)$ fields in (\ref{rescale}).  In the absence of
fine-tuning of the ratio of coupling constants we would expect $g'\sim
g$ in $D=6$, and hence $\td g' \sim \td g$ in $D=4$.

   Our discussion in this paper has focussed on the massless sector of
the dimensionally-reduced theory, but one can also study the massive
``Kaluza-Klein modes,'' at least in a linearised expansion around the
Minkowski vacuum.  It is easily seen that the mass scale $M$ for these modes
is given by
\be
\fft{M\, c}{\hbar}  
\sim \fft{1}{\cal R}\, e^{\ft14 \phi_0} \sim g\, e^{\ft12\phi_0} =
\td g\,,
\ee
which would therefore be of Planck scale if the physical 
Yang-Mills coupling constant were of order unity.  

    For example, if we set $\hat A_\1 = (2g)^{-1}\, \cos\theta\, d\varphi + A$,
then the contribution to the four-dimensional Lagrangian 
involving $A$, the massive boson associated with the (broken) 
$U(1)$ gauge symmetry of the six-dimensional theory, is found to be
\be
{\cal L}_A = -\ft12 e^{-\phi}\, {*F}\wedge F - 8 g^2\, {*A}\wedge A\,,
\ee
showing that $A$ is a Proca field with mass $ 4 g\, e^{\ft12\phi_0}\, 
(\hbar/c) = 4\td g\, (\hbar/c)$. 

    As another example, the tower of massive spin-2 fields will have
a mass spectrum given by the spectrum of the scalar Laplacian on the
compactifying 2-sphere.  Expanding around the Minkowski vacuum with 
$\phi=\phi_0$ we find that the masses are given by
\be
\fft{M_\ell \, c}{\hbar} = \fft{\sqrt{\ell\, (\ell+1)}}{{\cal R}}\,
    e^{\ft14\phi_0} =\sqrt{8\ell\, (\ell+1)}\, e^{\ft12\phi_0}\, g
= \sqrt{8\ell\, (\ell+1)}\,\td g \,.
\ee

     Our findings for the spectrum of massive modes differ in one
respect from those in \cite{quevedo}, where it was argued that there were two
distinct mass scales for Kaluza-Klein modes; a ``standard'' set of
states with scale $1/{\cal R}$, and a ``systematically light'' set with scale
$(1/{\cal R})\, e^{\ft14\phi_0}$.  In fact we find that {\it all}
massive modes have the latter scale, as is easily seen when one
expands the fields in harmonics around the vacuum solution: The
small fluctuations are governed, to leading order, by the
six-dimensional d'Alembertian $\hat{\square}_6$.  In the $\phi=\phi_0$
vacuum this takes the form $\hat{\square}_6 \sim e^{-\ft12\phi_0}\,
\square_4 + e^{\ft12\phi_0}\, \square_2$, and so with
four-dimensional (mass)$^2 \sim \square_4$, this implies a mass scale
$m\sim e^{\ft12\phi_0}\, g \sim \td g$, since $\square_2$ is the 
Laplacian on the $R_{mn}=8
g^2\, g_{mn}$ 2-sphere.  Thus there is a {\sl universal} mass-scale 
$\td g\sim (1/{\cal R})\, e^{\ft14\phi_0}$ for all massive modes in the
Kaluza-Klein spectrum.
Furthermore, all the Kaluza-Klein massive modes will be of Planck
scale and above, if the Yang-Mills coupling is taken to be of order of
magnitude unity.  

    We
have seen that owing to the presence of the dilaton field, the vacuum
energy scale is not set, as one might have supposed, by $1/{\cal
R}^4$, but rather by $M_{K}^4$, which is $(1/{\cal R}^4)\,
e^{\phi_0}$.  The fact that it is not ${\cal R}$ alone that occurs in
the expression for the vacuum energy is a reflection of the fact that
${\cal R}$ and $\phi_0$ separately have no intrinsic four-dimensional
physical significance; as we noted in section 2, $\phi_0$ can be
absorbed into a rescaling of the gauge coupling constant $g$.
However, an invariant statement about the four-dimensional bulk
Yang-Mills coupling constant for the Salam-Sezgin model is that
\be
g_{\rm YM} \sim \fft{M_K}{M_{\rm planck}}\,.
\ee
If the dimensionless Yang-Mills coupling constants are of
order unity, then the only scale in the problem is the Planck scale.

  In summary, in order to have a vacuum energy of the observed
(Hubble) magnitude, rather than $1/L_{\rm planck}^4$, one must choose
the $U(1)$ coupling constant to be about $10^{-31}$ in dimensionless
units.

   The model in \cite{quevedo2} introduced 3-branes at the north and
south poles of the 2-sphere, with tensions $T_3 = M_{\rm weak}^4$,
where $M_{\rm weak}\sim 100$ GeV, the weak scale.  The associated
deficit angles, in units of $2\pi$, are $\epsilon = T_3/(4 G_6)$,
where $G_6$ is the six-dimensional Newton constant, related to the
four-dimensional Newton constant $G$ by
\be
 G_6 = 4\pi\, G\, {\cal R}^2\, e^{\ft12\phi_0}\,.
\ee
Thus 
\be
\epsilon = \fft{\pi\, M_{\rm weak}^4\, {\cal R}^2\, e^{\ft12\phi_0}} 
               {M_{\rm planck}^2}\,.
\ee
By its very nature, $\epsilon$ cannot exceed 1.  

    The presence of the 3-branes alters the Dirac quantisation
condition.  Suppose that the solution is now supported by a Dirac
monopole configuration on $S^2$ that lies only partly in the original
$U(1)$ field $\hat F_\2$, with the remainder of the total contribution
required by the field equations supported within the additional
six-dimensional Yang-Mills sector $\hat F^I$;
\be
\hat F_\2 = \fft{\cos\beta}{2g}\, \Omega_\2\,,\qquad
T_I\, \hat F^I = T_0\, \fft{\sin\beta}{2g}\, \Omega_\2\,,
\ee
where $\beta$ is the ``mixing angle,'' and $T_0$ denotes the $U(1)$
generator within the Yang-Mills sector.  Of course if $\cos\beta \ne
1$, supersymmetry is broken.  Because the azimuthal coordinate $\varphi$ on
$S^2$ now has period $2\pi\, (1-\epsilon)$, the Dirac quantisation cannot
in general be simultaneously satisfied for both the $U(1)$ and Yang-Mills
groups.  Single-valuedness of the gravitino and the gauginos imply
\be
\cos\beta = \fft{N}{1-\epsilon}\,,\qquad 
\fft{g'\, \sin\beta}{g} = \fft{N'}{1-\epsilon}
\ee
respectively, where $N$ and $N'$ are integers. (The cases $\beta=0$ and
$\beta=\ft12\pi$ were given in \cite{quevedo2}, with the integers $N$ 
or $N'$ taken to be 1.)  In fact the first condition cannot
be satisfied for a positive-tension brane.  The second allows a
moderate-sized $g'$ for small $g$, if either the mixing angle $\beta$
is chosen to be very small, or else if the deficit-angle parameter
$\epsilon$ is chosen to be very close to its maximum value of unity.
Thus because
\be
\fft{g'}{g} = \fft{\td g'}{\td g}\,,
\ee
one could obtain a very small value for $\td g$, needed for the
suppression of the vacuum energy, whilst having $\td g'$ of order
unity, if $\epsilon$ were taken to be
equal to its maximum value 1 minus a quantity of order $10^{-31}$.
The resulting internal space, achieved by this choice, now
resembles a needle rather than a rugby ball, and has much in common
with the $S^1/Z_2$ orbifold of Horava and Witten \cite{horwit}.  This
suggests the intriguing possibility that in this {\sl maximal-tension
limit}, the six-dimensional Salam-Sezgin model approaches the
five-dimensional Horava-Witten models considered in \cite{ovrut}.  The
alternative possibility is to choose $\beta \sim
10^{-31}$.  Further investigation of these possibilities is clearly
worthwhile.

   Because the Kaluza-Klein scale is so small, the only viable
possibility would seem to be to get all the observed four-dimensional
gauge and matter fields from the 3-branes themselves, with all the
bulk gauge fields constituting a ``hidden sector'' \cite{quevedo2}.
In this scenario, there would be no direct phenomenological reason for
requiring the bulk gauge-field coupling constants $\td g$ and $\td g'$
to be of order unity, and so one might aim to achieve the
appropriately small value for the bulk vacuum energy $(1/{\cal R}^4)\,
e^{\phi_0}$ by choosing $\td g \sim 10^{-31}/L_{\rm planck}$.

     Issues of fine tuning or ``what is natural'' are complex, and
there seems to be no universally agreed notion or convention about
what they mean.  They are clearly dependent on the theory one considers,
and the choice of parameters specifying that theory.  In the present
case, the Salam-Sezgin model is completely specified by giving the
quantity $\td g = g\, e^{\ft12\phi_0}$, or equivalently, the
four-dimensional Yang-Mills coupling constant and the Planck mass.  In
terms of these, the choice $g_{\rm YM}\sim 10^{-31}$ could be said to be 
``fine tuned'' and ``unnatural.''
 
    This appearance of fine tuning might well change if the theory
were modified, for example by providing a potential that determined
the value of $\phi_0$, or if one were able to calculate quantities
such as the six-dimensional Newton constant, or six-dimensional
gauge-coupling constants, in terms of four-dimensional physical
quantities.  In the absence of a precise model that goes beyond the
level of supergravity, little more can be said at present.

\section{Conclusions}

    In this paper, we have shown that six-dimensional chiral $N=(1,0)$
gauged Einstein-Maxwell supergravity provides one of the rare examples
where a consistent dimensional reduction on a coset space is possible.
Specifically, we have shown that it admits a consistent Pauli
reduction on $S^2$, giving a four-dimensional theory, namely
$N=1$ supergravity coupled to an $SU(2)$ Yang-Mills vector multiplet
and a scalar multiplet.  Since, by definition, a consistent reduction
has the property that all solutions of the lower-dimensional equations
of motion provide solutions of the higher-dimensional equations of
motion, this reduction allows one to lift any solution of the
four-dimensional $N=1$ theory back to six dimensions.

   This dimensional reduction is of considerable interest in its own
right, since examples of consistent Pauli reductions are so few and
far between.  A further unusual feature, not seen in any of the other
known examples, is that only one scalar field (the breathing mode)
appears in the metric reduction ansatz.  In all other examples of
consistent Pauli reductions, such as the $S^7$ and $S^4$ reductions of
eleven-dimensional supergravity, the $S^5$ reduction of type IIB
supergravity, the $S^3$ reduction of the bosonic string, and the $S^2$
reduction of a certain Einstein-Maxwell-dilaton system, the inclusion
of $\ft12 n\, (n+3)$ scalars that parameterise inhomogeneous
distortions of the $S^n$ is necessary for consistency, since the
$SO(n+1)$ Yang-Mills fields act as sources for these scalars.  By
contrast, in the consistent $S^2$ reduction of the chiral
six-dimensional theory studied in this paper, the analogous source
terms turn out to arise with zero coefficient.\footnote{In fact
according to a linearised analysis of the entire spectrum of massless
and massive fields given in \cite{salsez}, these 5 scalars are members
of a massive supermultiplet, and so their omission from the consistent
Pauli reduction ansatz is likely to be obligatory rather than optional.}

   It is interesting also to note that had we been concerned only with
the reduction of the bosonic sector of the six-dimensional theory, a
broader class of consistent $S^2$ reductions would have been possible.
The reason for this is that in the purely bosonic sector, the only
place where the gauge-coupling constant $g$ appears in the
six-dimensional theory is in the scalar potential term in
(\ref{d6lag}).  One can then choose this to be distinct from the
constant $g$ appearing in the bosonic reduction ans\"atze
(\ref{metans}) and (\ref{fhphi}), without upsetting the consistency of
the reduction.  Thus if $g$ is relabelled as $g_0$ in the original
Lagrangian (\ref{d6lag}), while keeping the constant $g$ in
(\ref{metans}) and (\ref{fhphi}), we obtain a consistent reduction
that yields four-dimensional equations of motion derivable now from
\be
{\cal L} = R\, {*\oneone} - \ft12 {*d\phi}\wedge d\phi - \ft12
e^{-\phi}\, {*F^i}\wedge F^i -\ft12 e^{-2\phi}\, {*H_\3}\wedge H_\3
- 8({g_0}^2 - g^2)\, e^\phi\, {*\oneone}\,,
\label{4lagg}
\ee
where $F^i = dA^i + g\, \ep_{ijk}\, A^j\wedge A^k$ and $H_\3 = dB_\2 +
\ft12 \omega_\3$, where $d\omega_\3 = F^i\wedge F^i$.  This freedom to
construct a more general class of consistent reductions, yielding a
scalar potential in four dimensions, does not, however, extend to the
reduction of the full supergravity theory.  The gauge coupling
constant of the six-dimensional theory, which we have temporarily
relabelled as $g_0$ here, appears now in the gauge covariant
derivatives of all the fermionic fields; $(\hat\nabla_M -\im\, g_0\,
\hat A_M)\, \hat \chi$, {\it etc}.  One finds that the consistent reduction
of these fields, discussed in section \ref{fermsec}, is possible only
if $g_0=g$.  (For example, one obtains extra terms of the form $\im \,
(g_0-g)\, \mu_i\, A^i_\a\, \hat\ep$ in the transformation rule for
$\hat\psi_\a$, which would leave uncancelled $y$-dependence unless
$g_0=g$.)

   One question that is left unanswered by this work is whether the
six-dimensional chiral gauged supergravity can itself be obtained by
any consistent dimensional reduction from a yet higher dimension.
Various attempts have been made, but to date none has been successful.
Thus a string origin for the six-dimensional theory remains elusive.

\section*{Note Added}

    After the first version of this paper was submitted to the
archive, an M/string-theory origin for the Salam-Sezgin theory has
been found \cite{cvgipo}. Additionally, a new class of 3-brane
solutions in the Salam-Sezgin theory \cite{ggp} has been found, which
has some bearing on the issues of Dirac quantisation raised in section
5.

\section*{Acknowledgments} 
 
    We are grateful to Cliff Burgess, Rahmi G\"uven, Jim Liu, 
Fernando Quevedo, Ergin Sezgin and Paul Townsend for
discussions, and to the Feza Gursey Institute, Istanbul, for
hospitality during the completion of this work.  C.N.P. is grateful to
the Cambridge Relativity and Gravitation Group, for hospitality during
the earlier stages of this work.  We also thank Fernando Quevedo and Cliff
Burgess for discussions about an earlier version of section 5.

\appendix
\section{Curvature Calculations}

   For convenience, we present the results here for the spin
connection and the Ricci curvature for the following class of metrics:
\be
d\hat s^2 = e^{2\a\, \phi}\, ds^2 + e^{2\beta\, \phi}\, 
   g_{mn}\, (dy^m + \td g\, A^i\, K_i^m)(dy^n+ \td g\, A^j\, K_j^n)\,,
\ee
where $\a$ and $\beta$ are constants, $ds^2$ is the lower-dimensional
metric of dimension $d_x$, and $g_{mn}\, dy^m\, dy^n$ is the
undistorted ``internal'' metric, of dimension $d_y$.  We use the
orthonormal basis
\be
\hat e^\a = e^{\a\, \phi}\, e^\a\,,\qquad 
\hat e^a = e^{\beta\, \phi}\, (e^a + \td g\, A^i\, K_i^a)\,.
\ee
The Killing vectors $K_i^m$ on the internal space will be assumed to
satisfy the algebra $[K_i,K_j] = -f_{ij}{}^k\, K_k$.  

   The torsion-free spin connection is given by
\bea
\hat\omega_{\a\beta} &=&\omega_{\a\beta} + \a\,e^{-\a\, \phi}\,  
(\del_\beta\phi\, \hat e^\a - \del_\a\phi\, \hat e^\beta) - \ft12 \td g\, 
e^{(\beta-2\a)\phi}\, F^i_{\a\beta}\, K_i^a\, \hat e^a\,,\nn\\
\hat\omega_{\a b} &=& -\ft12\td g\, e^{(\beta-2\a)\phi}\,
F^i_{\a\beta}\, K_i^b\, \hat e^{\beta} -\beta\, e^{-\a\,\phi}\, 
\del_\a\phi\, \hat e^b  \,,\label{spincon}\\
\hat\omega_{ab} &=& \omega_{ab} - \td g\, e^{-\a\, \phi}\, A^i\,
\nabla_a K^i_b\, \hat e^\a\,,\nn
\eea
where $\omega_{\a\beta}$ and $\omega_{ab}$ are the spin connections for the
lower-dimensional spacetime and the undistorted internal space
respectively.  The Yang-Mills field strengths are defined by
\be
F^i = dA^i + \ft12 \td g\, f_{jk}{}^i\, A^j\wedge A^k\,.
\ee

   The orthonormal components of the Ricci tensor are given by
\bea
\hat R_{\a\beta} &=& e^{-2\a\, \phi}\, \Big[ R_{\a\beta} - \a\,
\square\phi\, \eta_{\a\beta} -(\a(d_x-2)+\beta
d_y)(\nabla_\a\nabla_\beta\phi +\a (\nabla\phi)^2\, \eta_{\a\beta})
\nn\\
&&
 + (\a^2(d_x-2)-\beta d_y(\beta-2\a))\nabla_\a\phi\, \nabla_\beta\phi
\Big] - \ft12\td g^2\, e^{(2\beta-4\a)\phi}\, F^i_{\a\gamma}\, 
F^j_{\beta}{}^\gamma\, K_i^a\, K_{j\, a}\,,\nn\\
\hat R_{\a b} &=& \ft12 \td g\, e^{(\beta-3\a)\phi}\,[ 
D^\beta F^i_{\a\beta} \, K_{i\, b} + [\a(d_x-4) + \beta
(d_y+2)]\, F^i_{\a\beta}\, K_{i\, b}\, \nabla^\beta\phi]\,,\label{ricci}\\
\hat R_{ab} &=& e^{-2\beta\, \phi}\, R_{ab} - \beta \,e^{-2\a\, \phi}\, 
[\square\phi + (\a(d_x-2)+ \beta d_y)(\nabla\phi)^2]\, \delta_{ab} 
 \nn\\
&& + \ft14\td g^2 \, e^{(2\beta-4\a)\phi}\, F^i_{\a\beta}\, F^{i\,
\a\beta}\, K_{i\, a}\, K_{j\, b}\,,\nn
\eea
where $R_{\a\beta}$ and $R_{ab}$ are the Ricci tensors of the
lower-dimensional spacetime and the undistorted internal space
respectively, and the derivative $D_\a$ is both spacetime and
Yang-Mills covariant.

   A canonical choice for the constants $\a$ and $\beta$ is to take
$\a\, (d_x-2) + \beta\, d_y=0$, since this ensures that the
higher-dimensional Einstein-Hilbert action will yield a normal
Einstein-Hilbert term in the lower dimension, with no power of 
$e^\phi$ multiplying it.  It can bee seen from (\ref{ricci}) that this
choice leads to considerable simplifications.

   Our specific case in this paper has $d_x=4$, $d_y=2$, $\a=-\beta=
\ft14$ and $\td g= 2g$, and so we shall have
\bea
\hat R_{\a\beta} &=& e^{-\ft12 \phi}\, \Big[ R_{\a\beta} - \ft14\,
\square\phi\, \eta_{\a\beta} -\ft14 \nabla_\a\phi\, \nabla_\beta\phi
\Big] - 2 g^2\, e^{-\ft32\phi}\, F^i_{\a\gamma}\, 
F^j_{\beta}{}^\gamma\, K_i^a\, K_{j\, a}\,,\nn\\
\hat R_{\a b} &=& g\, 
D^\beta(e^{-\phi}\,  F^i_{\a\beta} )\, K_{i\, b}\,,\label{ricci2}\\
\hat R_{ab} &=& e^{\ft12\, \phi}\, R_{ab} +\ft14 \,e^{-\ft12 \phi}\, 
\square\phi\, \delta_{ab} 
  + g^2 \, e^{-\ft32\phi}\, F^i_{\a\beta}\, F^{i\,
\a\beta}\, K_{i\, a}\, K_{j\, b}\,,\nn
\eea


\begin{thebibliography}{99}

\bm{kaluza} T. Kaluza, {\it On the problem of unity in physics},
Sitzunber. Preuss. Akad. Wiss. Berlin. Math. Phys. {\bf K1}, 966 (1921).

\bm{dewitt} B.S. DeWitt, in {\it Relativity, groups and topology},
Les Houches 1963 (Gordon and Breach, 1964).


\bm{strau} N. Straumann, {\it On Pauli's invention of non-Abelian
Kaluza-Klein theory in 1953}, gr-qc/0012054.

\bm{oraf} L. O'Raifeartaigh and N. Straumann, {\it Early history of
gauge theories and Kaluza-Klein theories, with a glance at recent
developments}, hep-ph/9810524.

\bm{dewnic} B. de Wit and H. Nicolai, {\it The consistency of the 
$S^7$ truncation in $D = 11$ supergravity},
Nucl. Phys. {\bf B281}, 211 (1987).

\bm{nasvamvan} H. Nastase, D. Vaman and P. van Nieuwenhuizen,
{\it Consistency of the AdS$_7\times S^4$ reduction and the origin of  
self-duality in odd dimensions},
Nucl. Phys.  {\bf B581}, 179 (2000), hep-th/9911238.

\bm{s5red}
M. Cveti\v c, H. L\"u, C.N. Pope, A. Sadrzadeh and T.A. Tran,
{\it Consistent $SO(6)$ reduction of type IIB supergravity on $S^5$},
Nucl. Phys. {\bf B586}, 275 (2000), hep-th/0003103.

\bm{spherered}
M. Cveti\v c, H. L\"u and C.N. Pope,
{\it Consistent Kaluza-Klein sphere reductions},
Phys. Rev. {\bf D62}, 064028 (2000), hep-th/0003286.

\bm{cvgilupo} M. Cveti\v c, G.W. Gibbons, H. L\"u and C.N. Pope,
{\it Consistent group and coset reductions of the bosonic string},
hep-th/0306043.

\bm{salsez}
A. Salam and E. Sezgin,
{\it Chiral compactification on (Minkowski)$\times S^2$ of $N=2$ 
Einstein-Maxwell supergravity in six-dimensions},
Phys. Lett. {\bf B147}, 47 (1984).

\bm{nishsez} H. Nishino and E. Sezgin,
{\it Matter and gauge couplings of $N=2$ supergravity in six dimensions},
Phys. Lett. {\bf B144}, 187 (1984).

\bibitem{quevedo}
Y. Aghababaie, C.P. Burgess, S.L. Parameswaran and F. Quevedo,
{\it SUSY breaking and moduli stabilization from fluxes in gauged 6D  
supergravity},
JHEP {\bf 0303}, 032 (2003), hep-th/0212091.

\bm{quevedo2}
Y. Aghababaie, C.P. Burgess, S.L. Parameswaran and F. Quevedo,
{\it Towards a naturally small cosmological constant from branes in 6D  
supergravity}, hep-th/0304256.

\bm{horwit}  P. Horava and E. Witten, {\sl Heterotic and type I string dynamics
from eleven dimensions}, Nucl. Phys. {\bf B460} (1996) 506.

\bm{ovrut} A. Lukas, B.A. Ovrut, K.S. Stelle and D. Waldram, {\sl Heterotic
M-theory in five dimensions}, Nucl. Phys. {\bf B552} (1999) 246.

\bm{cvgipo} M. Cveti\v c, G.W. Gibbons and C.N. Pope, 
{\it A  string and M-theory origin for the Salam-Sezgin model},
Nucl.\ Phys. {\bf B677}, 164 (2004), hep-th/0308026.

\bm{ggp}G.W. Gibbons, R. G\"uven and C.N. Pope,
{\it 3-Branes and uniqueness of the Salam-Sezgin vacuum}, hep-th/0307238.


\end{thebibliography}
\end{document}